\documentclass{pnastwo}
\usepackage{graphicx}
\usepackage{hyperref}
\usepackage[numbers]{natbib}
\begin{document}
\title{Scaling and Hierarchy in Urban Economies} \author{Cosma Rohilla
  Shalizi\affil{1}{Statistics Department, Carnegie Mellon University, 5000
    Forbes Avenue, Pittsburgh, PA 1513 USA}\affil{2}{Santa Fe Institute, 1399
    Hyde Park Road, Santa Fe, NM 87501 USA}}
\maketitle

\begin{article}

\begin{abstract}
  In several recent publications, Bettencourt, West and collaborators claim
  that properties of cities such as gross economic production, personal income,
  numbers of patents filed, number of crimes committed, etc., show super-linear
  power-scaling with total population, while measures of resource use show
  sub-linear power-law scaling.  Re-analysis of the gross economic production
  and personal income for cities in the United States, however, shows that the
  data cannot distinguish between power laws and other functional forms,
  including logarithmic growth, and that size predicts relatively little of the
  variation between cities.  The striking appearance of scaling in previous
  work is largely artifact of using extensive quantities (city-wide totals)
  rather than intensive ones (per-capita rates).  The remaining dependence of
  productivity on city size is explained by concentration of specialist service
  industries, with high value-added per worker, in larger cities, in accordance
  with the long-standing economic notion of the ``hierarchy of central
  places''.
\end{abstract}
\keywords{Applied statistics | Model comparison | Urban economics | Urban scaling | Central place hierarchy | Non-parametric smoothing}
\abbreviations{MSA, metropolitan statistical area; GMP, gross metropolitan product; BEA, Bureau of Economic Analysis; RMS, root mean square}

\section{Introduction}

\dropcap{R}ecent dramatic advances in explaining metabolic scaling relations in
biology by the properties of optimal transport networks
\cite{West-Brown-Enquist-origin-of-scaling,Brown-West-scaling-in-bio} suggest
the possibility of examining social assemblages, especially cities, in similar
terms.  In a well-known series of papers, Bettencourt, West and collaborators
\cite{Bettencout-et-al-growth-innovation-scaling,Bettencourt-et-al-urban-scaling-and-its-deviations}
claim that many social and economic properties of cities --- gross economic
production, total personal income, number of patents filed, number of people
employed in ``supercreative'' \cite{Florida-rise-of-creative-class}
occupations, number of crimes committed, etc. --- grow as super-linear powers
of population size, while measures of total resource use grow as sub-linear
powers.  These two claims imply that per capita output grows as a positive
power of population, while per capita resource use shrinks as a negative power.
If reliable and precise scaling laws of this type exist, they would be of
considerable importance for both science and policy
\cite{Bettencourt-West-unified-theory-of-urban-living}\footnote{Bettencourt and
  West summarize their claims regarding this ``unified theory of urban living''
  \cite{Bettencourt-West-unified-theory-of-urban-living} thus: ``We have
  recently shown that these general trends [to cities] can be expressed as
  simple mathematical laws''; ``Our work shows that, despite appearances,
  cities are approximately scaled versions of one another \ldots: New York and
  Tokyo are, to a surprising and predictable degree, nonlinearly scaled-up
  versions of San Francisco in California or Nagoya in Japan. These
  extraordinary regularities open a window on underlying mechanism, dynamics
  and structure common to all cities''; ``Surprisingly, size is the major
  determinant of most characteristics of a city; history, geography and design
  have secondary roles''.}.

Reasonable arguments from long-standing principles of economic geography would
lead one to expect that larger cities would have higher economic output per
capita, through a combination of the benefits to firms in related industries
clustering together (``agglomeration economies''), and the tendency of firms
and specialists with large increasing returns to scale to be located high in
the ``hierarchy of central places''.  (For reviews of these concepts, including
historical notes, mathematical models and empirical evidence, see Refs.\
\cite{Krugman-self-organizing,Henderson-ZMS-Venables,Fujita-Krugman-Venables}.)
These arguments would carry over to producing technologically useful knowledge
and to ``supercreative'' services as well.  However, these economic
considerations do not point to either a particular functional form for the
growth of per-capita output with population, or suggest that it should be very
strong.  Moreover, these theories do not look at individual cities as isolated
monads, as scaling arguments do, but rather rely on there being assemblages of
multiple cities (and rural areas), coupled by common economic processes, and
assuming distinct roles in those processes through a history of mutual
interaction and combined and uneven development.

The purpose of this note is to argue that, at least for the United States,
while there is indeed a tendency for per-capita economic output to rise with
population, power-law scaling predicts the data no better than many other
functional forms, and worse than some others.  Furthermore, the impressive
appearance of scaling displayed in Refs.\
\cite{Bettencout-et-al-growth-innovation-scaling,Bettencourt-et-al-urban-scaling-and-its-deviations}
is largely an aggregation artifact, arising from looking at extensive
(city-wide) variables rather than intensive (per-capita) ones.  The actual
ability of city size to predict economic output, no matter what functional form
is used, is quite modest.  These conclusions hold whether economic output is
measured by gross metropolitan product or by total personal income.  If we
control for metropolitan areas' varying concentration of industrial sectors, we
find that the remaining scaling with population is negligible, and much of the
variance across cities is predicted by the extent to which they host specialist
service providers with strongly increasing returns, as predicted by the idea of
the hierarchy of central places.

I begin by re-analyzing the gross metropolitan product data, showing that
scaling is far weaker than it seemed in Refs.\
\cite{Bettencout-et-al-growth-innovation-scaling,Bettencourt-et-al-urban-scaling-and-its-deviations}. I
then \ref{sec:pace-of-life} re-analyzes the data on walking speed, originating
in Ref.\ \cite{Bornstein-Bornstein-pace-of-life} and presented in Ref.\
\cite{Bettencout-et-al-growth-innovation-scaling}, which makes the problems
with the scaling analysis very clear.  Per-capita productivity is better
predicted by how much a city depends on industrial sectors which indicate a
high position in the hierarchy of specialist service provision.  This actually
eliminates any significant role for scaling with size.  The conclusions
summarize the scientific import of the data analyses.

The supplemental information shows that (i) scaling also fails for personal
income, (ii) the hypothesis of power-law scaling cannot be saved by positing a
mixture of distinct scaling relations, and that (iii) {\em contra} Ref.\
\cite{Bettencourt-et-al-urban-scaling-and-its-deviations}, neither a Gaussian
nor a Laplace distribution is a good fit to the deviations from the power-law
scaling relations.

All calculations were done using R \cite{R}, version 2.12.  Code for reproducing
figures and analyses is included in the supplemental information.

\section{Models}

\paragraph{Scaling Models} 
Ref.\ \cite{Bettencout-et-al-growth-innovation-scaling} reported a power-law
scaling between the population of cities in the United States and their
economic output.  To be precise, the units of analysis are ``metropolitan
statistical areas'' (MSAs) as defined by the official statistical
agencies\footnote{To quote Ref.\ \cite{McCormick-Panek-Rodriguez-gmp-for-2009},
  MSAs are ``standardized county-based areas that have at least one urbanized
  area with a population of 50,000 or more plus adjacent territory that has a
  high degree of social and economic integration with the core, as measured by
  commuting ties.''}.  The measure of economic output is the gross domestic
product for each metropolitan area (``gross metropolitan product'' or GMP), as
calculated by the U.S. Bureau of Economic Analysis
(\url{http://www.bea.gov/regional/gdpmetro/}), which is supposed to be the sum
of all ``incomes earned by labor and capital and the costs incurred in the
production of goods and services'' in the metropolitan area
\cite{Assenova-Broda-Taylor-gdp-by-state}\footnote{A word on the BEA's
  procedure is in order \cite{Panek-Baumgardner-McCormick-gdp-by-msa}. The BEA
  estimates gross products for each industry for each state, and conducts
  surveys to estimate what fraction of each industry's state-wide earnings is
  located in each metropolitan area.  Multiplying these ratios by the
  state-wide gross products, and summing over industries, gives the gross
  metropolitan product.  The BEA provides no estimates of measurement
  uncertainty for these numbers.}.  Ref.\
\cite{Bettencout-et-al-growth-innovation-scaling} analyzed data for 2006,
deflated to constant 2001 dollars, and I will do likewise; the 2008 and 2004
data are not much different.

Ref.\ \cite{Bettencout-et-al-growth-innovation-scaling} propose that output
scales as a power of population, $Y \propto N^b$.  This is connected to the
data via the linear regression model
\begin{equation}
\ln{Y} = \ln{c} + b\ln{N} + \epsilon ~,
\label{eqn:power-law-model}
\end{equation}
with $\epsilon$ being a mean-zero noise term.  For later comparisons, it will
be convenient to denote this by $Y \sim c N^b$.

There is a simple test of the model which has not, so far as I know, been
applied before.  If production does scale as some power of population, $Y \sim
cN^b$, then per-capita production $Y/N \equiv y$ should also scale,
\begin{equation}
 y \sim c N^{b-1} ~,
\label{eqn:per-capita-power-law-model}
\end{equation}
and vice versa.  As shown below, this transformation drastically changes the
apparent fit of the power-law scaling model.

It is worth noting that there is no theoretical reason to expect a power-law
scaling relation of the form of Eq.\ \ref{eqn:power-law-model} for urban
economies (while there are such reasons for biological scaling
\citep{West-Brown-Enquist-origin-of-scaling,Brown-West-scaling-in-bio}).
Accordingly, and unlike Ref.\
\cite{Bettencout-et-al-growth-innovation-scaling}, I also consider a
logarithmic scaling model,
\begin{equation}
y \sim r \ln{N/k}
\label{eqn:logarithmic-scaling}
\end{equation}
a logistic scaling model,
\begin{equation}
\ln{y} \sim d_1 + d_2 \frac{e^{(N-d_3)/d_4}}{1+e^{(N - d_3)/d_4}}
\label{eqn:logistic-scaling}
\end{equation}
and finally a non-parametric scaling relation,
\begin{equation}
\ln{y} \sim s(\ln{N})
\label{eqn:smooth-scaling}
\end{equation}
with $s$ an unknown smooth function, to be determined by the data.  By
comparing multiple scaling models, including the fully data-driven Eq.\
\ref{eqn:smooth-scaling}, we can see to what extent the data actually provide
evidence for particular functional form, or indeed for any strong scaling
relation at all.

Notice that Eq.\ \ref{eqn:per-capita-power-law-model} implies per-capita output
grows with population without limit, and with constant marginal elasticity (the
``$15\%$ rule'' of \cite{Bettencourt-West-unified-theory-of-urban-living});
according to Eq.\ \ref{eqn:logarithmic-scaling} the growth is unlimited, but
slows as population grows; and according to Eq.\ \ref{eqn:logistic-scaling},
per-capita output is asymptotically constant in population.

\paragraph{Urban Hierarchy} 
An alternative to size scaling is hierarchical structure.  The ``hierarchy of
central places'', introduced by L{\"o}sch and Christaller in the 1930s, has
become a corner-stone of urban economic geography.  In outline, the idea is
that developed economies contain many specialized goods, and especially
services, that the mass of consumers need only rarely (such as the services of
a surgeon), or indirectly (such as the services of a professor of surgery, or a
maker of surgical instruments).  The provision of such services has
comparatively high fixed costs (the time needed to train a surgeon) but low
marginal costs (the time needed to perform an operation), leading to increasing
returns to scale.  It thus becomes economically efficient for these specialists
to locate in central places, where their fixed costs can be distributed over
large consumer bases, and the more specialized they are, the more centrally
located they need to be, and the larger the customer base they require.  This
logic leads to the formation of a hierarchy of market centers and cities, in
which increasingly specialized skills, with (as it were) increasingly
increasing returns, can be had, and so predicts positive associations between
the population of urban centers, the concentration of specialist skills within
them, and (owing to increasing returns) their per-capita economic output.  Good
reviews of the theory, including historical citations and connections to modern
economic models of increasing returns, may be found in Refs.\
\cite{Krugman-self-organizing,Fujita-Krugman-Venables}.

Fortunately, the BEA also provides estimates of the shares of gross
metropolitan products attributable to different industrial sectors, some of
which correspond to the specializations emphasized in central place theory.  I
specifically consider ``Information, Communication, and Technology (ICT)'',
``Financial activities'', ``Professional and technical services'' and
``Management of companies and enterprises'' (industry codes 106, 102, 58 and
62, respectively)\footnote{The BEA withholds the GMP-contribution figures for
  some industry-MSA combinations, when the sector is so concentrated in that
  city that releasing the number would provide consequential business
  information about specific firms.  I have fit the model discussed below for
  the 133 cities with complete data in the four selected sectors.
  Experimenting with various forms of imputation for the missing data did not
  materially change the results.}.  Writing the proportions of gross
metropolitan product deriving from each of these sectors as $x_1$ through
$x_4$, the level of per-capita production can be predicted by a log-additive
model \cite{Buja-Hastie-Tibshirani-additive-models} which incorporates
power-law scaling with city size:
\begin{equation}
\ln{y} = \ln{c} + b\ln{N} + \sum_{j=1}^{4}{f_j(x_j)} + \epsilon ~,
\label{eqn:additive-model}
\end{equation}
where each of the ``partial response'' functions $f_j$ summarizes the
contribution of the $j^{\mathrm{th}}$ industrial sector.  For comparison with
the power-law scaling model (Eq.\ \ref{eqn:power-law-model}), I have
constrained the partial response function for size to be logarithmic; the other
partial response functions can be nonlinear, though they must be smooth.

\section{Statistical Methods}

Power-law scaling relations, like Eq.\ \ref{eqn:power-law-model}, were
estimated through ordinary least squares, i.e., minimizing
$n^{-1}\sum_{i=1}^{n}{(\ln{Y_i} - \ln{c} - b\ln{N_i})^2}$, where the index $i$
runs over metropolitan areas, of which there are $n=366$.  As is well known
\cite{Wasserman-all-of-stats}, this is a consistent estimator of regression
parameters for transformed regressions, even when Gaussian noise assumptions
are violated, though the nominal values of standard errors and confidence
intervals cannot be trusted\footnote{The analogous procedure for fitting
  power-law {\em distributions} is not reliable, due to differences between
  regression and density estimation.}.  The nonlinear but parametric models
(Eq.\ \ref{eqn:logarithmic-scaling} and \ref{eqn:logistic-scaling}) were fit by
nonlinear least squares.

The non-parametric size scaling relation, Eq.\ \ref{eqn:smooth-scaling}, was
fit by means of a smoothing spline \cite{tEoSL-2nd,Wahba-spline-models} on the
logged data.  That is, the estimated spline is the function $s$ minimizing
\begin{equation}
n^{-1}\sum_{i}{(\ln{y_i} - s(\ln{N_i}))^2} +  \lambda\int{(s^{\prime\prime}(x))^2dx}
\end{equation}
with the smoothness penalty $\lambda > 0$ chosen by cross-validation.
Smoothing splines of this type are universal approximating functions, and
picking the penalty by cross-validation controls the risk of over-fitting
non-generalizing aspects of the data --- see Ref.\ \cite{Wahba-spline-models}
for details.\footnote{A smoothing spline fit to the un-transformed data was
  similar, but visually somewhat more jagged.}

Finally, the additive model (Eq.\ \ref{eqn:additive-model}) was estimated by
combining spline smoothing for the non-parametric partial response functions
$f_j$, and an iterative ``back-fitting'' procedure
\cite{Buja-Hastie-Tibshirani-additive-models}.  (I used the \texttt{mgcv}
library \cite{Wood-mgcv-2004}.)  This adjusts for the correlations between
industrial sectors, and between city size and industrial sectors, so that each
estimated partial response function is, as far as possible, the unique additive
contribution of that variable to economic output.

\section{Results}

\subsection{Weakness of Scaling in Gross Metropolitan Products}
\label{sec:no-scaling}

\begin{figure}
\includegraphics[width=\columnwidth]{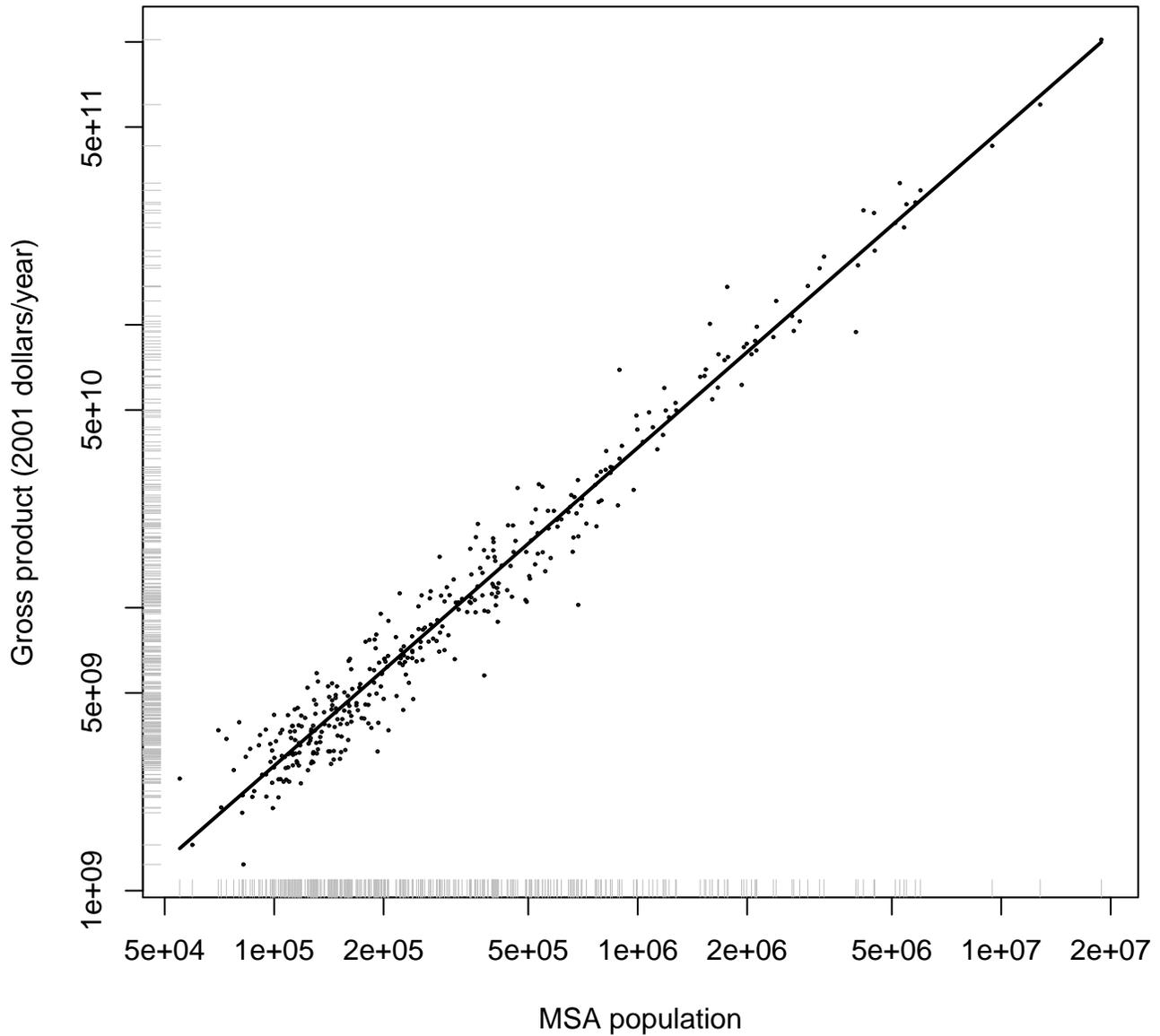}
\caption{Horizontal axis: population of the 366 US metropolitan statistical
  areas in 2006, log scale; vertical axis, 2006 gross product of each MSA, in
  constant 2001 dollars, log scale.  (In all figures, grey inner ticks on axes
  mark observed values.)  Solid line: ordinary least squares
  regression of log gross metropolitan product on log population, i.e., the
  regression $Y \sim c N^b$, with estimated exponent $\hat{b} = 1.12$.}
\label{fig:gmp-2006}
\end{figure}

Fitting Eq.\ \ref{eqn:power-law-model} by least squares, I estimate $b$ to be
$1.12$, in agreement with Ref.\
\cite{Bettencout-et-al-growth-innovation-scaling}, with a 95\% bootstrap
confidence interval \cite{Wasserman-all-of-stats} of $(1.10, 1.15)$.  Figure
\ref{fig:gmp-2006} shows the data and the fitted trend, with both axis plotted
on a logarithmic scale, so that a power law relationship appears as a straight
line.  The root-mean-squared (RMS) error for predicting $\ln{Y}$ is $0.23$, and
the ``coefficient of determination'' $R^2$ is $0.96$, i.e., the fitted values
retain 96\% of the variance in the actual data.

\begin{figure}
\includegraphics[width=\columnwidth]{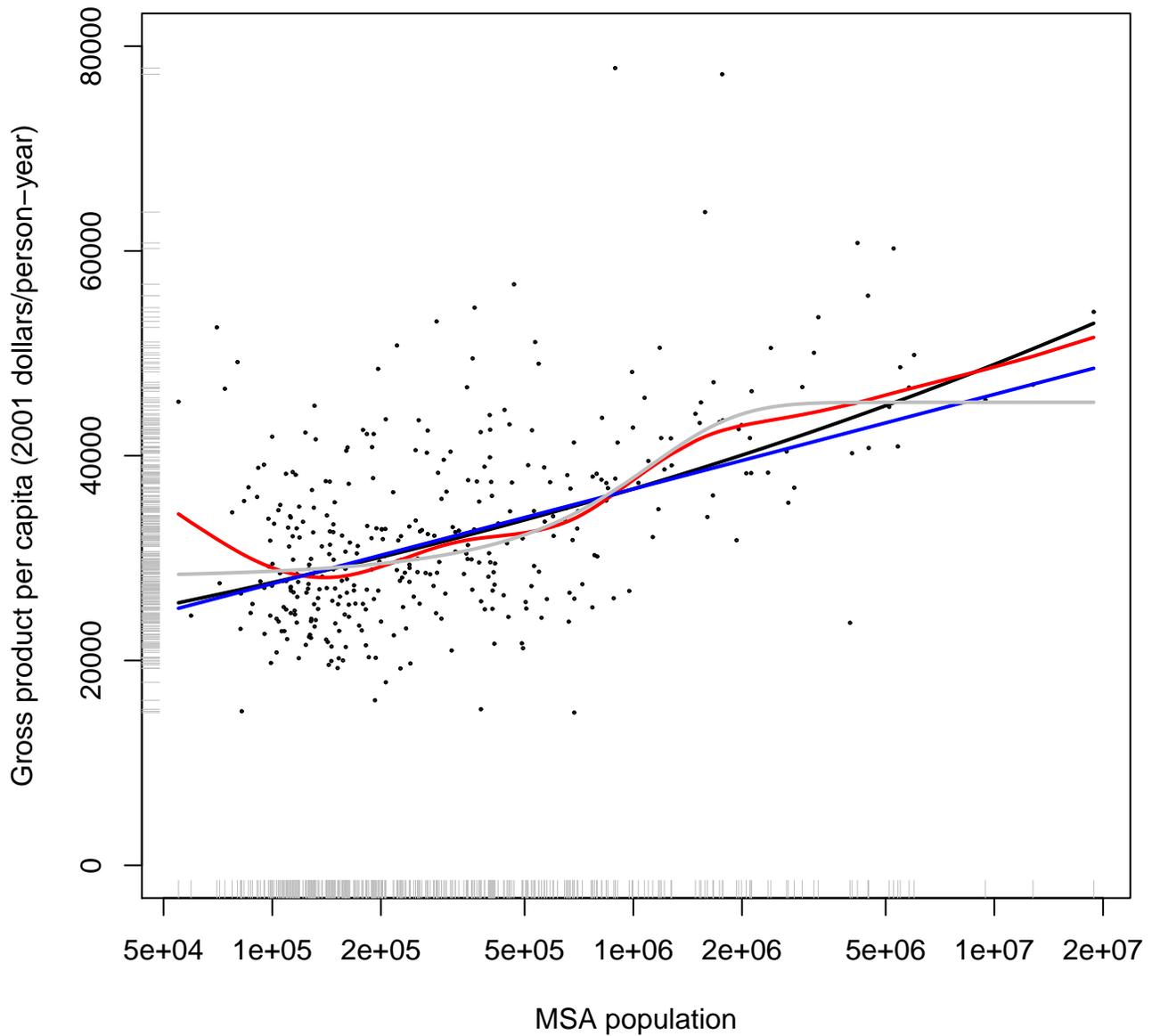}
\caption{Horizontal axis: population, as in Figure \ref{fig:gmp-2006}, log
  scale.  Vertical axis: gross product per capita, but on a linear and not a
  logarithmic scale.  The two largest values are $7.8\times{10}^4$
  dollars/person-year (in Bridgeport-Stamford-Norwalk, CT, a center for hedge
  funds and other financial firms) and $7.7\times{10}^4$ dollars/person-year
  (in San Jose-Sunnyvale-Santa Clara, CA, i.e., Silicon Valley), and the
  smallest are $1.5\times{10}^4$ dollars/person-year (in
  McAllen-Edinburg-Mission, TX and Palm Coast, FL).  Black line: fitted
  power-law scaling relation.  Blue line: fitted logarithmic scaling
  relationship.  Grey line: logistic scaling.  Red line: smoothing spline
  fitted to the logged data.}
\label{fig:pcgmp-vs-pop-2006}
\end{figure}

Visually, this looks like reasonable data collapse.  Plotting the per-capita
values $y$ however, as in Figure \ref{fig:pcgmp-vs-pop-2006}, reveals a very
different picture, though the two should be logically equivalent under the
power-law model.

Figure \ref{fig:pcgmp-vs-pop-2006} shows a trend curve for the the power-law
scaling implied by Ref.\ \cite{Bettencout-et-al-growth-innovation-scaling}.
(The exponent estimated for $y$ is $0.12$, matching that estimated for $Y$, as
it must.)  The figure also shows the fitted logarithmic scaling relationship
(Eq.\ \ref{eqn:logarithmic-scaling}), which is extremely close to the power law
over the range of the data, the logistic scaling relationship (Eq.\
\ref{eqn:logistic-scaling}), and the non-parametric smoothing spline,
corresponding to the relationship $y \sim e^{s(\ln{N})}$.  Note that the latter
curve is not even monotonically increasing in $N$.

While the curves in Figure \ref{fig:pcgmp-vs-pop-2006} correspond to very
different modeling assumptions --- the differences between the implications of
power-law and logistic scaling are perhaps especially striking --- they all
account for the data about equally well, or rather, equally poorly, because
most of the variation in per-capita production is, in fact, unrelated to
population.  (Note that the vertical axis is linear, not logarithmic.) The RMS
errors for $\ln{y}$ of the power law, of logarithmic scaling and of logistic
scaling are, respectively, $0.232$, $0.234$ and $0.229$, while that of the
spline is $0.225$.  They would predict $y$, for a randomly chosen city, to
within $\pm 26.1$, $\pm 26.3$, $\pm 25.7$ and $\pm 25.3$ percent, respectively.
Predicting the same value of $y$ for all cities, however, has an RMS error of
$0.27$, a margin of $\pm 30\%$.  Thus the $R^2$ values are, respectively,
$0.24$, $0.23$, $0.26$ and $0.29$.  On the linear scale, i.e., in terms of
dollars per person-year $y$, the RMS errors of the power law, logarithmic,
logistic and spline curves are, respectively, $7.9\times{10}^3$,
$7.9\times{10}^3$, $7.8\times{10}^3$ and $7.7\times{10}^3$, as compared to
$9.2\times{10}^3$ for predicting the mean for all cities.\footnote{All these
  measures of error are calculated on the same data used to fit the models,
  exaggerating the models' predictive powers.  However, using six-fold
  cross-validation to approximate the out-of-sample risk gives RMS errors of
  $0.234$ for the power law, $0.236$ for logarithmic scaling, $0.232$ for
  logistic scaling, and $0.231$ for the spline.  The differences are small, but
  bootstrapping shows they are significant at the 5\% level (at least).}  In
other words, even allowing for quite arbitrary functional forms, city size does
not predict economic output very well.

The similarity of the RMS errors, and indeed of the curves, arises in part from
the limited range of $y$.  The difference between the largest and smallest
per-capita products ($6.3\times{10}^4$ dollars/person-year) is ``only'' a
factor of 5.2, i.e., not even one order of magnitude.  This is too small, with
only 366 observations, to distinguish among competing functional forms for the
trend, while still being quite consequential in human and economic terms.
Per-capita production is simply not very strongly related to population.

Taking any per-capita (intensive) quantity which is statistically independent
of population, and looking at the corresponding aggregate (extensive)
quantities will yield a scaling exponent close to one.  The overwhelming
majority of the apparent fit of the scaling relationship in Figure
\ref{fig:gmp-2006} is just such an artifact of aggregation.  This can be shown
in three different ways: by algebra; by extrapolating the different per-capita
functional forms back to city-wide totals; and by simulation.

Algebraically, suppose that $y$ was statistically independent of $N$.  Then
$\ln{Y} = \ln{y} + \ln{N}$ would be the sum of two independent random
variables, so its variance would be the sum of their variances.  The $R^2$ of a
linear regression of $\ln{Y}$ on $\ln{N}$, with the slope constrained to be 1,
would be $\frac{\mathrm{Var}[\ln{N}]}{\mathrm{Var}[\ln{N}] +
  \mathrm{Var}[\ln{y}]}$, which with this data comes to $0.94$
\cite{Wasserman-all-of-stats}.  That is, {\em even if} intensive, per-capita
output was completely independent of city size, fitting a power-law scaling
model to the aggregated data would capture $94\%$ of the variance in the
extensive, total output.  The actual $R^2$, on the other hand, is only
$0.96$.\footnote{Examples like this are why regression textbooks advise against
  using $R^2$ to check goodness of fit
  \cite{Birnbaum-devil-rides-again,Anderson-Shanteau-weak-inference,Berk-on-regression}.}

\begin{figure}
\includegraphics[width=\columnwidth]{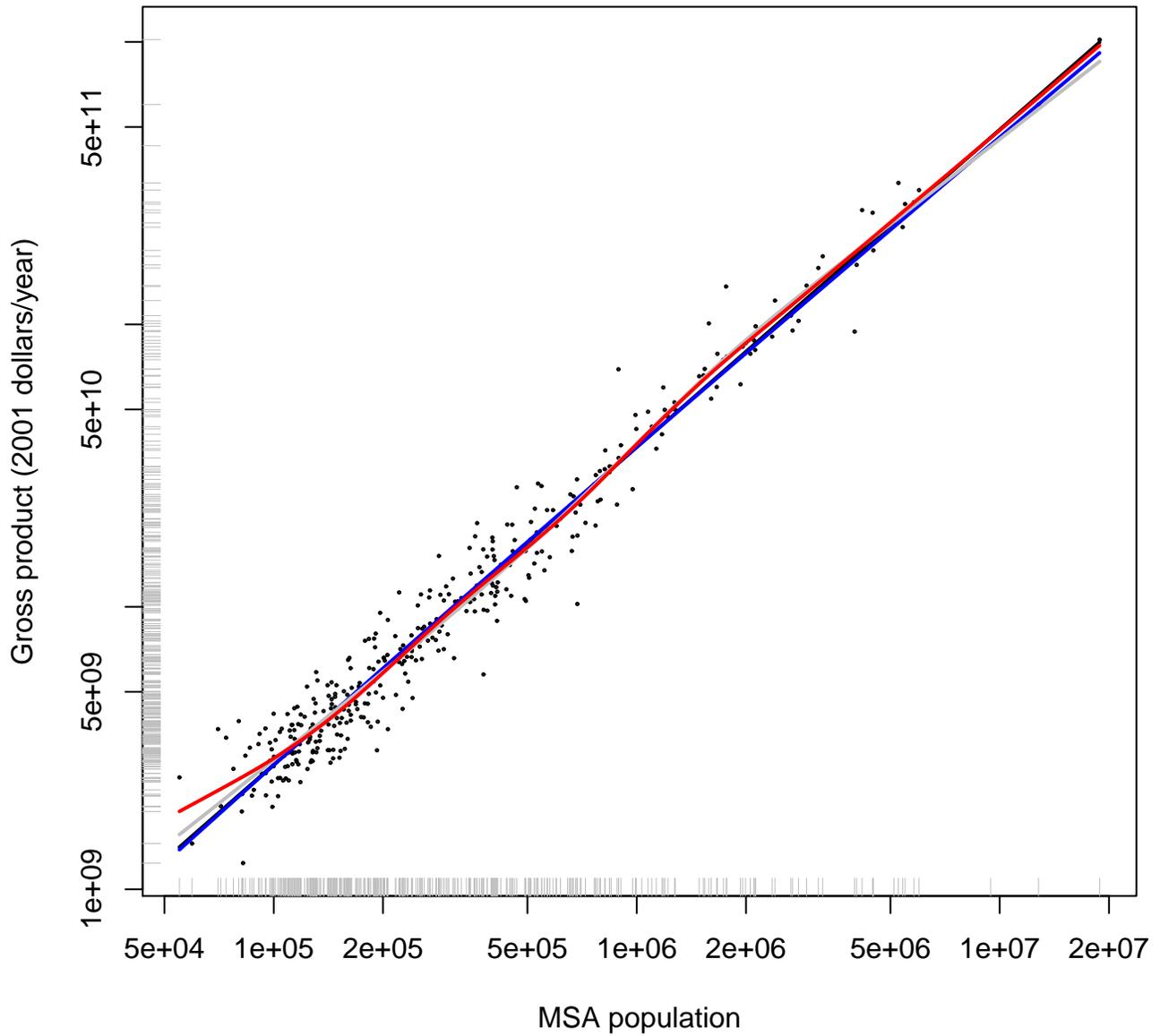}
\caption{As in Figure \ref{fig:gmp-2006}, but with the addition of curves
  showing the scaling relations from Figure \ref{fig:pcgmp-vs-pop-2006},
  extrapolated to aggregate rather than per-capita values.  These are visually
  all but indistinguishable.}
\label{fig:gmp2006-ols-vs-slope-1}
\end{figure}

Figure \ref{fig:gmp2006-ols-vs-slope-1} shows the same data and scaling curve
as Figure \ref{fig:gmp-2006}, but three additional trend lines.  These are the
logarithmic, logistic and spline fits to the per-capita data (from Figure
\ref{fig:pcgmp-vs-pop-2006}) extrapolated back to the implied aggregated values
$Y$.  These are, visually, almost indistinguishable from the fitted power law;
all have $R^2 = 0.96$.

\begin{figure}
\includegraphics[width=\columnwidth]{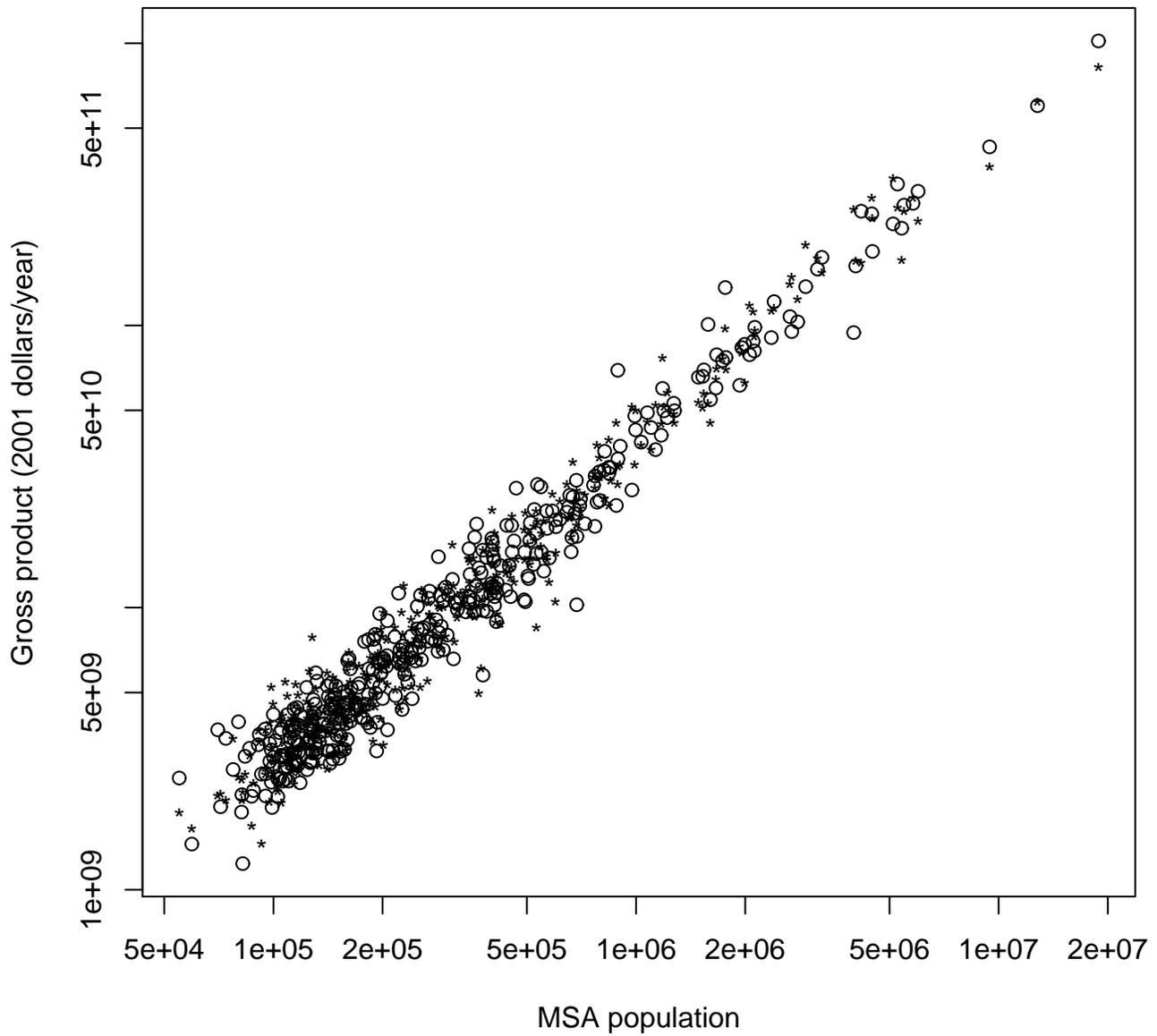}
\caption{Axes: as in Figure \ref{fig:gmp-2006} and
  \ref{fig:gmp2006-ols-vs-slope-1}.  Circles: Actual values.  Stars: simulated
  values, with per-capita production figures drawn from the logistic (not
  power-law) scaling model.}
\label{fig:gmp-vs-loggrowth-surrogate}
\end{figure}

Figure \ref{fig:gmp-vs-loggrowth-surrogate} demonstrates in a different way
that the data do not support the idea of power-law scaling.  The circles in the
figure show the actual data values.  The stars, by contrast, are surrogate data
simulated from the fitted logistic scaling model, with the actual population
sizes.  The surrogate per-capita output values $\tilde{y}$ were set equal to
the fitted values under the model of Eq.\ \ref{eqn:logistic-scaling}, and then
randomly perturbed according to the empirical distribution of deviations from
that model.  The figure plots the surrogate aggregate products $\tilde{y}N$,
which look very much like the data.

If a power-law scaling relation is fit to the surrogate data from the
logistic-scaling regression, then, averaging over many simulations, the median
scaling exponent is $1.12$, with 95\% of the estimates falling between $1.10$
and $1.15$, and the median $R^2$ of the power-law was $0.96$.  Recall that the
estimate for the actual data was $1.12$, with a 95\% confidence interval of
$(1.10, 1.15)$, and $R^2=0.96$.

The RMS error for $\ln{y}$ on the real data is very slightly lower for the
logistic model ($0.229$) than for the power law ($0.232$).  The difference is
minute, but is, in fact, statistically significant: when repeating both fits on
surrogate data simulated from the power law, gaps of this size or larger occur
only $\approx 1\%$ of the time.  Not too much should be read into this,
however, owing to the small magnitude of the difference, the large errors
around both regression curves, and the comparatively small number of
observations.  Reliably discriminating between the two models simply requires
more information (in the sense of \cite{Kullback-info-theory-and-stats}) than
the data provides: either much smaller fluctuations of $y$ around the
regression curve, or many more data points.

To sum up these results, the appearance of a strong, super-linear relationship
between gross production $Y$ and population $N$ is mostly driven by production
growing in proportion to population --- that is, linearly.  Per-capita
production $y$ does not have a strong scaling relationship of {\em any} form
with $N$, and the data are unable to distinguish between different functional
forms for such trends as there are.  The same is true of personal income (SI,
section 1).  Lacking ready access to the data sets on patents, crime,
infrastructure and resource consumption which Ref.\
\cite{Bettencout-et-al-growth-innovation-scaling} analyzed in the same way as
economic output and personal income, I cannot say whether the reported scaling
relations for those aggregate variables suffer from the same problem.

\subsection{Superiority of the Urban-Hierarchy Model}

To address the question of why there is a weak and noisy tendency for
per-capita output to rise with population, I turn to the log-additive model,
Eq.\ \ref{eqn:additive-model}.  Fitting to the data yields the partial response
functions shown in Figure \ref{fig:partial-response-functions}.  As expected
from the urban-hierarchy argument, all four of the partial response functions
are monotonically increasing, so that rising shares of those industries predict
increasing per capita production.  Very notably, however, the estimated
power-law scaling exponent is actually negative, $-2.6\times{10}^{-3}$, but
statistically indistinguishable from zero (standard error
$2.8\times{10}^{-2}$).  That is, in the log-additive model, controlling for
these four industrial sectors makes population effectively irrelevant for
predicting urban productivity.  Indeed, dropping population from the model
altogether produces no appreciable difference in the fit.  At least at the
level of expectation values, controlling for these four industrial sectors
``screens off'' the effects of city size on per-capita production.

Statistically, there is no question that the log-additive model predicts better
than the simple scaling model.  The RMS error of the former, on the log scale,
is $0.218$, corresponding to an $R^2$ of $38.8\%$, and an accuracy of $\pm
24\%$ or $\$6.8\times{10}^3$, better than any model based on size alone.  The
log-additive model is a more flexible specification, and so over-fitting to the
data is an issue, but this can be addressed by cross-validation, which directly
measures the ability of a model to extrapolate from one part of the data to
another \cite{tEoSL-2nd}.  The cross-validated mean squared error of the
log-additive model is $0.053$, while that of the pure power law is $0.067$,
clearly showing that the extra complexity of the former is being used to
capture genuinely predictive patterns, and not merely to memorize the training
data\footnote{Dropping population size $N$ from the log-additive model
  altogether actually improves the cross-validation score, very slightly, to
  $0.052$.}.

The simple log-additive model is unlikely to be a fully adequate predictor of
systematic differences in urban productivity.  If nothing else, these four
coarse-grained industrial sectors were selected merely for convenience, as
approximate indicators of position in the urban hierarchy, and presumably could
be improved.  Moreover, the model does not even try to represent the
interactive processes which lead cities to have the industrial mixes that they
do.  In reality, these industries can be so concentrated towards the largest
cities, at the top of the hierarchy (e.g., New York), and away from lower-rank
cities (e.g., San Francisco, Peoria), only because all these cities are part of
a single national, and even international, division of labor
\cite{Fujita-Krugman-Venables}.

\subsection{``The Pace of Life''}
\label{sec:pace-of-life}

\begin{figure}
\includegraphics[width=\columnwidth]{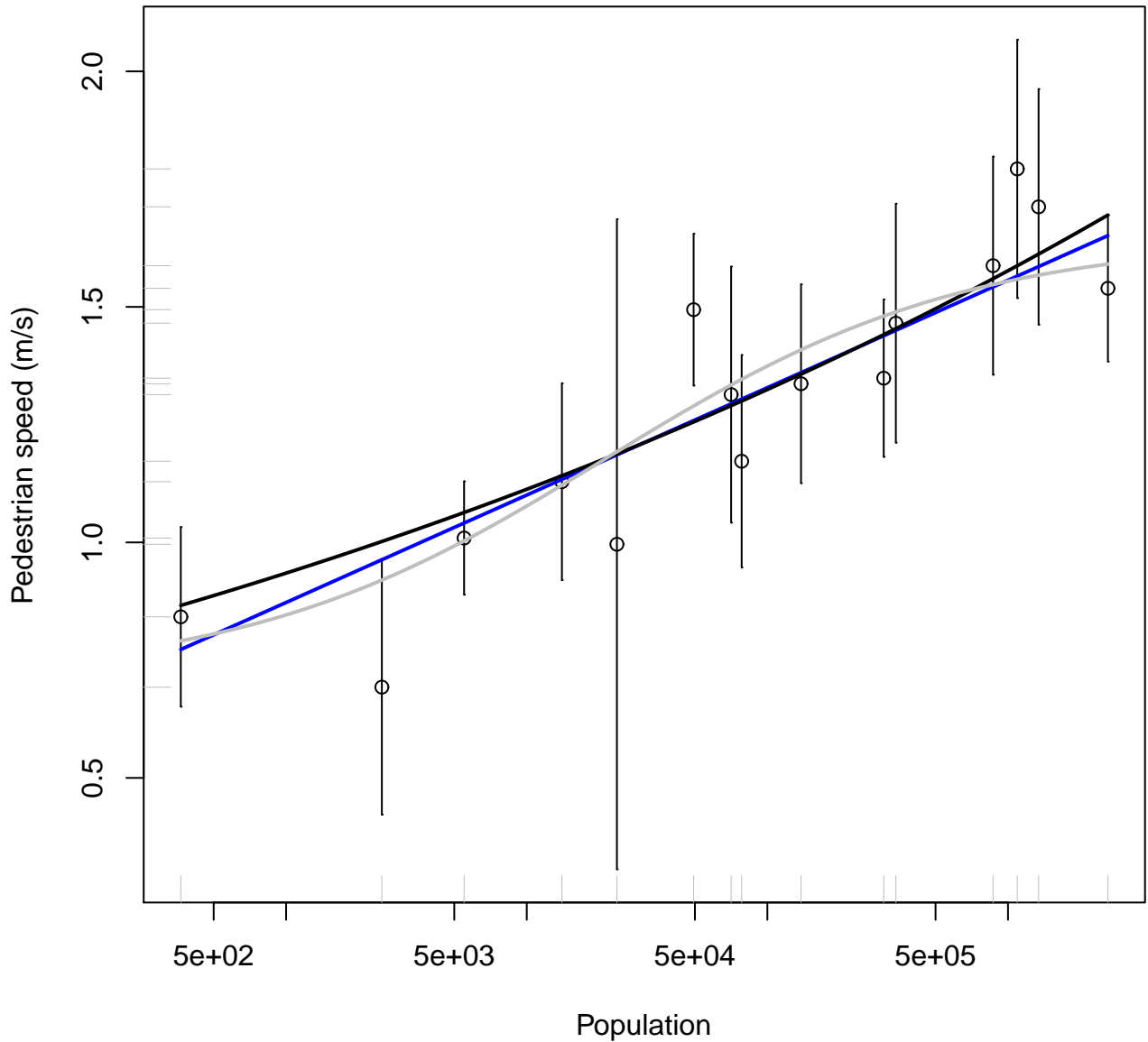}
\caption{Horizontal axis: city population, logarithmic scale.  Vertical axis:
  estimated pedestrian speed in meters/second, plus or minus one standard
  deviation, linear scale.  Blue line: the regression $v \sim r\ln{N/k}$, as
  proposed by Ref.\ \cite{Bornstein-Bornstein-pace-of-life}.  Black line: the
  regression $v \sim c N^b$, as proposed by Ref.\
  \cite{Bettencout-et-al-growth-innovation-scaling}.  Grey line: logistic
  scaling.  (Data from Ref.\ \cite{Bornstein-Bornstein-pace-of-life}, who
  report the mean and standard deviation of the time taken to walk 50 feet =
  15.2 meters; I calculated standard deviations by propagation of error.)}
\label{fig:pace-of-life}
\end{figure}

\begin{figure*}
\includegraphics{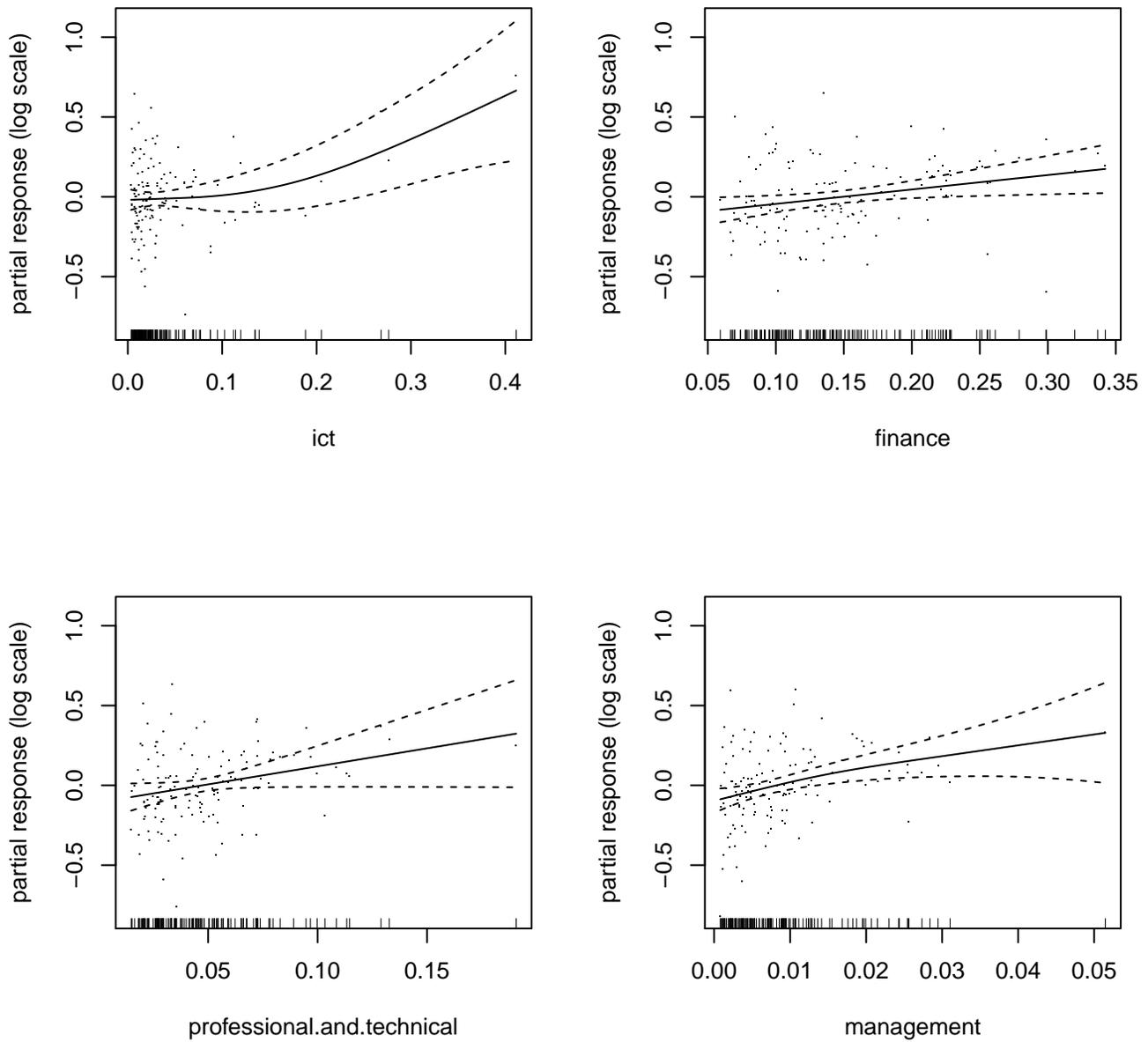}
  \caption{Partial response functions for the log-additive model (Eq.\
    \ref{eqn:additive-model}).  Horizontal axes indicate the fraction of each
    metropolitan area's gross product derived from each industry, while the
    vertical axis shows the predicted logarithmic increase, or decrease, to per
    capita output, relative to the baseline of the mean over all cities.  Solid
    curves are the main estimate, with dashed curves at $\pm 2$ standard errors
    in the partial response function.  Dots show ``partial residuals'', the
    difference between actual $\ln{y}$ values and those predicted by the model
    including all the other variables.}
\label{fig:partial-response-functions}
\end{figure*}

A further claim of Ref.\ \cite{Bettencout-et-al-growth-innovation-scaling} is
that the speed at which people walk grows as a positive power of the number of
people in a city.  The source given for this is Ref.\
\cite{Bornstein-Bornstein-pace-of-life}, a two-page letter to {\em Nature} in
1976.  The authors of Ref.\ \cite{Bornstein-Bornstein-pace-of-life} went to 15
cities, towns and villages, picked locations and individuals which seemed to
them to be comparable, and timed how long it took them to walk fifty feet
($15.2$ meters).  Such unsystematic data, however intriguing, is too weak to
support substantial scientific conclusions.  Nonetheless, it is instructive to
examine it, as in Figure \ref{fig:pace-of-life}.

The original plot (Figure 1 in Ref.\ \cite{Bornstein-Bornstein-pace-of-life})
showed population on a log scale, and speed on a linear scale, as in Figure
\ref{fig:pace-of-life}.  The linear regression, for this transformation of the
data, corresponds to assuming that speed grows logarithmically with population,
$v \sim r\ln{N/k}$.  Figure 2a in Ref.\
\cite{Bettencout-et-al-growth-innovation-scaling} re-plots the same data, but
with the vertical axis on a logarithmic scale, so the linear regression assumes
speed grows as a power of population, $v \sim c N^b$.  (Neither figure included
error bars, though Bornstein and Bornstein give the standard deviations in
their caption.)  As can be seen from Figure \ref{fig:pace-of-life}, both of
these regressions, along with logistic scaling, are very similar in this data,
while they embody very different assumptions, and at most one can be right.

The explanation for this apparent paradox is that the range of reported walking
speeds is small, from $0.7$ m/s to $1.8$ m/s, and if $|x| \ll 1$, then
$\ln{1+x} \approx x$.  Observed over a narrow range, then, logarithmic and
power law scaling simply are very similar, and hard to distinguish.  This is
also why the the power-law and logarithmic fits to per-capita production in
Figure \ref{fig:pcgmp-vs-pop-2006} were so close.

\section{Discussion}
\label{sec:conclusion}

Neither gross metropolitan product nor personal income scales with population
size for U.S.\ metropolitan areas.  The appearance of scaling in Refs.\
\cite{Bettencout-et-al-growth-innovation-scaling,Bettencourt-et-al-urban-scaling-and-its-deviations}
is an artifact of inappropriately looking at extensive variables (city-wide
totals) rather than intensive ones (per-capita values).  Scaling is also
unpersuasive for walking speed.  I was not able to examine the other variables
claimed to show scaling in Refs.\
\cite{Bettencout-et-al-growth-innovation-scaling,Bettencourt-et-al-urban-scaling-and-its-deviations},
but, as they were all extensive variables, the analyses reported there would be
subject to the same aggregation artifacts.  It is also possible that cities in
the contemporary United States are anomalous, and that scaling of income and
economic output holds elsewhere.

It is evident from Figures \ref{fig:pcgmp-vs-pop-2006} (and Supplemental Figure
S1) that there is a weak tendency for per-capita output and income to rise with
population, though the relationship is simply too loose to qualify as a scaling
law.  (Arguably, the real trend in those figures is for the {\em minimum}
per-capita output to rise with population, though I would press this point.)
Qualitatively, this is what one would expect from well-established findings of
economic geography \cite{Henderson-ZMS-Venables}.  The data do not really
support any stronger {\em quantitative} statement.  In particular, asserting
any specific functional form, such as a power law, goes far beyond the what the
data can support.  Nor is there any theory, supported on independent grounds,
which predicts a specific functional form.  Accordingly, extrapolations based
on such claims (e.g., the finite-time singularity in the model for city growth
in \cite{Bettencout-et-al-growth-innovation-scaling}) are speculative at best.
The amplitude of fluctuations around the trend lines are, in any case, so large
that predictions based on size alone can have very little utility.

By taking account of the shares of just a few industries in the gross
metropolitan product, we can obtain much better predictions of the level of
per-capita production.  In this statistical model, summarized in Eq.\
\ref{eqn:additive-model}, population plays no significant direct role in
predicting per capita economic output, and could in fact be profitably ignored.
Rather, the industrial sectors used are chosen as signs of where metropolitan
areas stand in the urban hierarchy, which is also related, of course, to size.
One could interpret this as the mechanism by which size scaling happens (to the
extent that it does), but this would imply that an exogenous increase in a
city's population would automatically shift its industrial pattern, which is
implausible.  Indeed, the whole scaling picture for cities seems to rest on an
oddly monadic, interaction-free view of metropolitan areas.  The logic of
central place theory, in contrast, relies on cities being part of an
interactive assemblage, coupled by processes of production, distribution and
exchange.  This not only seems more plausible, but also better matches the
evidence at hand.

As Refs.\
\cite{Bettencout-et-al-growth-innovation-scaling,Bettencourt-et-al-urban-scaling-and-its-deviations,Bettencourt-West-unified-theory-of-urban-living}
have stressed, developing a sound scientific understanding of cities should be
a priority for an increasingly urban species.  In seeking such understanding,
it is a sound strategy to begin with simple hypotheses, and to reject them in
favor of more complicated ones only as they prove unable to explain the data.
This is not because the truth is more likely to be simple, in some metaphysical
sense, but because this strategy leads us to the truth faster and more reliably
than ones which invoke needless complexities
\cite{Kelly-simplicity-truth-probability}.  The elegant hypothesis of power-law
scaling marked a step forward in our understanding of cities, but it is now
time to leave it behind.

\appendix{}

\section{Personal Income}
\label{sec:personal-income}

\begin{figure}
\includegraphics[width=\columnwidth]{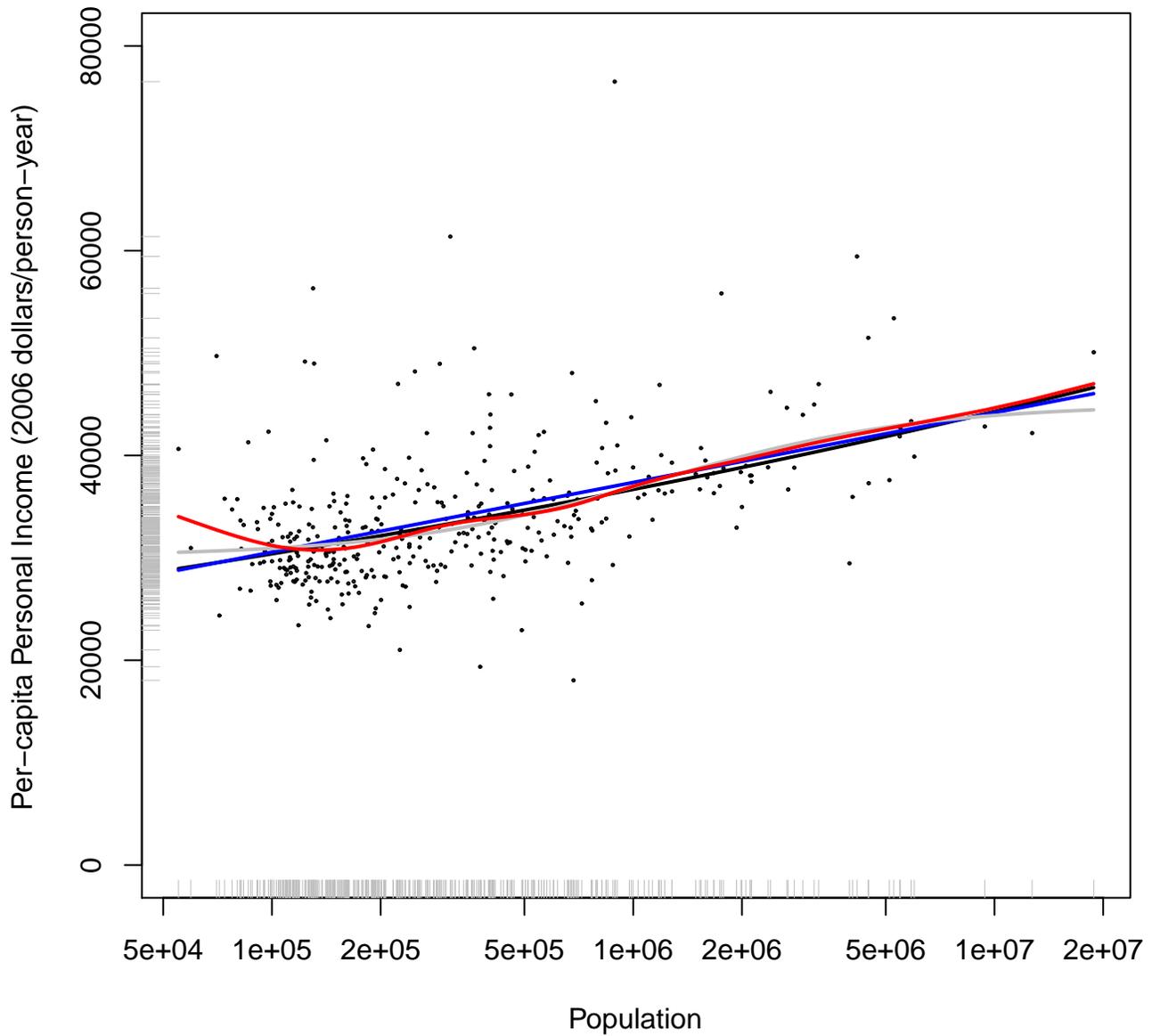}
\caption{Personal income per capita versus population, 2006.  Horizontal axis:
  population of MSAs (log scale).  Vertical axis: personal income per capita,
  in nominal 2006 dollars (linear scale).  Black line: power-law scaling curve
  (estimated exponent $0.082$).  Blue line: logarithmic scaling curve.  Grey
  line: logistic scaling curve.  Red line: spline fit to logged data.}
\label{fig:personal-income-per-capita}
\end{figure}

The BEA also makes available estimates of personal income by metropolitan area,
a variable closely related to, but not quite the same as, the gross
metropolitan product.  (See \url{http://www.bea.gov/regional/reis/} for
definitions, estimation techniques, and data.)  Ref.\
\cite{Bettencourt-et-al-urban-scaling-and-its-deviations} reports that total
personal income $L$ also scales as a power of population, implying per capita
personal income $L/N \equiv l$ should scale likewise.  Figure
\ref{fig:personal-income-per-capita} plots $l$ versus $N$, with the
best-fitting power law, logarithmic relationship, and spline.

Once again, the appearance of power-law scaling in the aggregate variable is
not supported by examination of the per-capita values.  The RMS error, on the
log scale, of predicting a constant per capita income over all cities is
$0.179$, while the RMS errors of the power-law, logarithmic and logistic
scaling relations are $0.157$, $0.158$ and $0.156$, and that of the spline
$0.154$.  Even the spline thus has an $R^2$ of only $0.26$.  Repeating the
procedures of Figures 3 and 4 from the main text yields similar results.  Thus,
personal income also fails to display non-trivial scaling with population.

\section{Mixtures of Scaling Relations}
\label{sec:mixture-of-regressions}

Recall that the posited scaling relation is $y \sim cN^b$.  As shown above,
this does not fit the data, at least not assuming, following Ref.\
\cite{Bettencout-et-al-growth-innovation-scaling}, that both parameters, the
scaling exponent $b$ and the pre-factor $c$, are the same for all cities.  A
natural way to try to reconcile the data with the model would be to modify the
latter, allowing $c$ to depend on the {\em type} of the city.  The rationale
for such a regression would be that there are several different kinds of
cities, and that city type shifts the over-all level of production up or down,
but, once that is factored out, all cities scale with size in the same way.
This common scaling exponent would not, naturally, be the same as the one
estimated from the pooled data.

Formally, we introduce a latent variable $Z$ for each city, treated as a
discrete random variable independent of $N$, and consider the statistical model
$y \sim c_Z N^b$.  This leads to a ``mixture-of-regressions'' or ``latent-class
regression'' model, which can be fit by the expectation-maximization algorithm
\cite{Benaglia-et-al-mixtools}.  Such fitting would lead not only to estimates
of $b$ and the pre-factors $c_z$, but also to the probability that each city
belonged to each of the different city types or mixture components,
categorizing cities inductively from the data.

To investigate this, I fit mixture-of-regression models to the data from Figure
2 in the main text, varying the number of mixture components from 1 to 10,
using the software of Ref.\ \cite{Benaglia-et-al-mixtools}.\footnote{For
  computational reasons, it is easier to fit the more general specification in
  which the scaling exponent is also allowed to vary, $y \sim c_Z N^{b_Z}$.
  (Sharing a parameter across the regressions complicates the maximization step
  of the expectation-maximization algorithm.)  If the constant-exponent model
  is right, the estimated exponents for each mixture component should agree to
  within statistical precision.}  To determine the correct number of mixture
components, I used both Schwarz's ``Bayesian'' information criterion and
cross-validation, which are both known to be consistent for such mixture
problems, unlike the Akaike information criterion, which over-fits
\cite{Claeskens-Hjort-model-selection}.  Both BIC and cross-validation strongly
favored {\em one} mixture component, meaning that the fit to the data is not
actually improved by allowing for multiple scaling curves.

This does not completely rule out the $y \sim c_Z N^b$ model, as only 366
observations may not have enough information to simultaneously induce
appropriate categories and fit scaling relations.  An alternative would be to
expand the information available, by defining the categorical variable $Z$ in
terms of measurable attributes of cities other than $N$ and $y$, such as
geographic location or the mix of industries.  (See Ref.\
\cite{Gelman-Hill-data-analysis} on such variable-intercept, constant-slope
regressions with known categories.)  Success with such models hinges on
selecting categories to represent important features of the data-generating
process, a task I must leave to other inquirers.

Assuming that such a statistical model works, there would still be the question
of its interpretation.  Whether one would judge such a model to really show
scaling in urban assemblages would depend on how much importance one gives, on
the one hand, to a common scaling exponent, and on the other to most of the fit
coming from the un-modeled differences across city types. 

\section{Residuals}
\label{sec:residuals}

Ref.\ \cite{Bettencourt-et-al-urban-scaling-and-its-deviations} proposes
ranking cities not by their per capita values of quantities like economic
production or patents or crime, but by the deviation, positive or negative,
from the scaling relationship, i.e., by the residuals of the trend lines.  (It
does not compare this to ranking by per capita values.  The Spearman rank
correlation between the two variables is $0.87$ for GMP and $0.83$ for personal
income.)  They consider both a Gaussian distribution for the residuals, i.e., a
probability density $f(x) \propto e^{-x^2/2\sigma^2}$, and a Laplace
distribution, $f(x) \propto e^{-\lambda|x|}$, and claim that both fit very
well.

\begin{figure}
\includegraphics[width=\columnwidth]{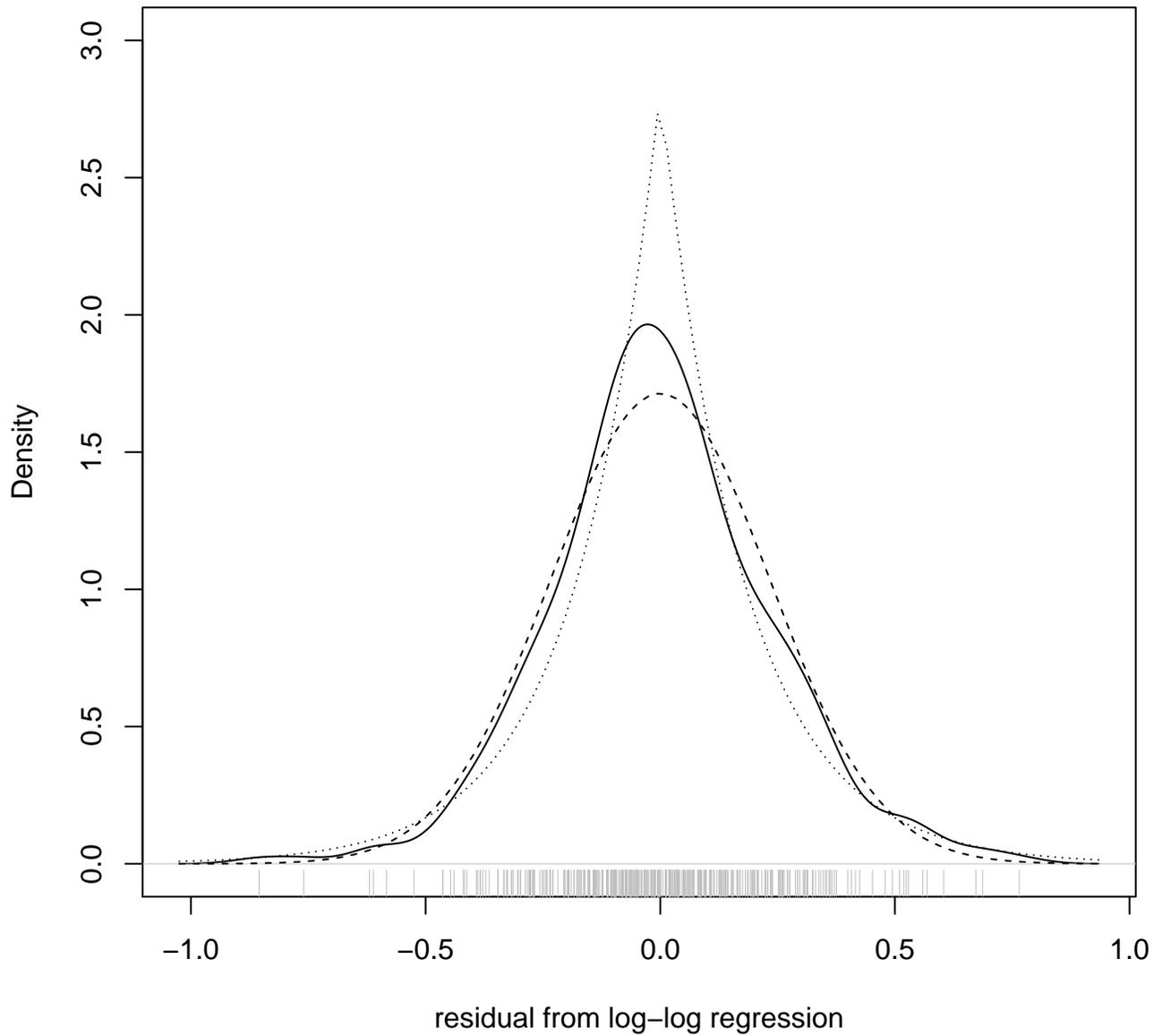}
\caption{Horizontal axis: magnitude of residuals from power-law scaling of
  gross metropolitan product on population, i.e., from regressing $\ln{Y}$ on
  $\ln{N}$.  Vertical axis: probability density of the residual distribution.
  Solid line: Nonparametric kernel density estimate (Gaussian kernel, default
  bandwidth choice --- see Ref.\ \cite{Simonoff-on-smoothing}).  Dashed line:
  maximum likelihood Gaussian fit to residuals.  Dotted line: maximum
  likelihood Laplace (double-exponential) fit to residuals.}
\label{fig:distribution-of-residuals}
\end{figure}

Figure \ref{fig:distribution-of-residuals} shows the situation for GMP.
Visually, neither distribution matches the residuals well.  Quantitatively,
goodness-of-fit can be checked by ``data-driven smooth tests''
\cite{WCMKallenberg-Ledwina-data-driven-smooth-tests}, which transform their
inputs so that they will be uniform if and only if the postulated distribution
holds, and then measure departures from uniformity (coefficients from expanding
the transformed empirical distribution in a series of orthogonal polynomials).
Such tests reject both the Gaussian and the Laplace distribution with high
confidence ($p$-values of $1\times{10}^{-3}$ and $8\times{10}^{-3}$,
respectively, calculated using code provided by Ref.\ \cite{ddst}).

\begin{figure}
  \includegraphics[width=\columnwidth]{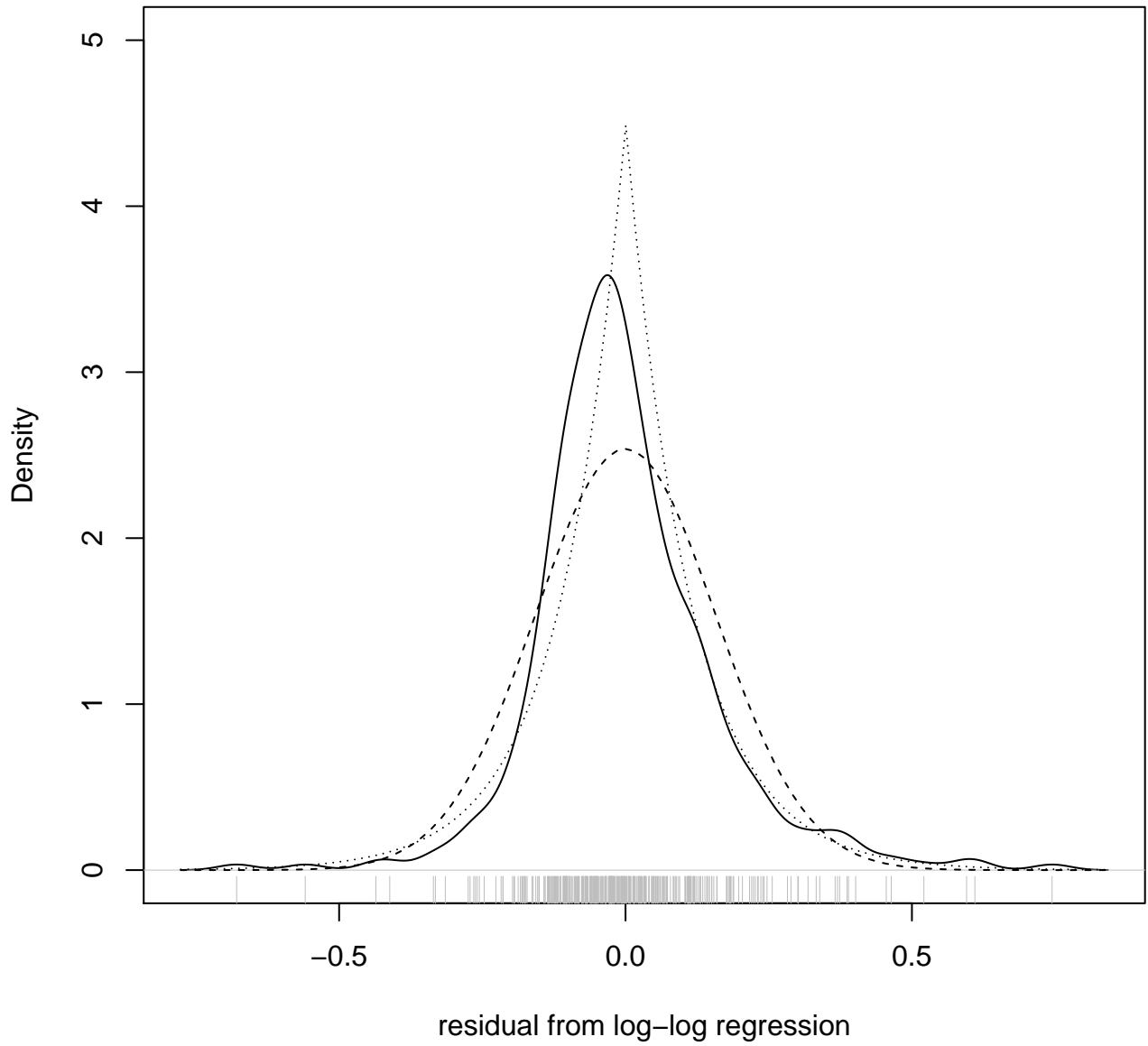}
  \caption{As in Figure \ref{fig:distribution-of-residuals}, but showing the
    deviations of personal income from power-law scaling.}
\label{fig:residuals-for-income}
\end{figure}

Results for personal income are similar (Figure
\ref{fig:residuals-for-income}).  The Gaussian distribution can be rejected
with high confidence ($p < {10}^{-4}$).  While the data do not rule out the
Laplace distribution in the same way ($p=0.27$), the limited power of the test
at the comparatively small sample size means that there is not strong evidence
in its favor either.  (See Ref.\ \cite{Mayo-Cox-frequentist} on the evidential
interpretation of significance tests.)

\begin{acknowledgments}
Thanks to Jordan Ellenberg for valuable questions,
and to grants from the NIH (\# 2 R01 NS047493) and the Institute for New
Economic Thinking for partial support.
\end{acknowledgments}

\bibliographystyle{unsrtnat}
\bibliography{locusts}
\end{article}

%Figure 1

%Figure 2

%Figure 3

%Figure 4

%Figure 5

%Figure 6

%Figure 7

%Figure 8

%Figure 9

\end{document}